\documentclass[10pt]{article}

\usepackage[english]{babel}

\usepackage{graphics,color}
\usepackage{newlfont,epsfig}
\usepackage{amssymb}
\begin{document}
\date{}
\begin{center}
{\Large\bf Evolution of a quantum harmonic oscillator coupled to a minimal thermal environment}
\end{center}
\begin{center}
{\normalsize A. Vidiella-Barranco \footnote{vidiella@ifi.unicamp.br}}
\end{center}
\begin{center}
{\normalsize{ Instituto de F\'\i sica ``Gleb Wataghin'' - Universidade Estadual de Campinas}}\\
{\normalsize{ 13083-859   Campinas  SP  Brazil}}\\
\end{center}
\begin{abstract}
In this paper it is studied the influence of a minimal thermal environment on the dynamics of a quantum harmonic oscillator 
(labelled $A$), prepared in a coherent state. The environment itself consists of a second oscillator (labelled $B$), 
initially in a thermal state. Two types of interaction Hamiltonians are considered, and the time-evolution of the reduced 
density operator of oscillator $A$ is compared to the one obtained from the usual master equation approach, i.e., 
assuming that oscillator $A$ is coupled to a large reservoir. An analysis of the linear entropy evolution of oscillator 
$A$ shows that simplified models may be able to describe important features related to the phenomenon of decoherence.
\end{abstract}

\setcounter{footnote}{0}
\section{Introduction}

The coupling of a quantum system to an environment normally leads to the degradation of its non-classical properties. 
Usually, the environment is modeled by a large number of quantum systems (the reservoir) e.g., a collection of
independent harmonic oscillators. However analytical solutions of models involving large reservoirs are virtually 
impossible to obtain, and approximations are generally necessary. For instance, by assuming a weak system-reservoir 
coupling, it is possible, via perturbative methods, to derive evolution equations (master equations) 
for the reduced density operator\footnote{As one is interested in the evolution of the quantum system itself, a 
partial trace is taken over the environment variables.} of the system of interest 
\cite{louisell73}. This approach, based on the assumption of the existence of a large reservoir, naturally leads to 
irreversible dynamics of the system variables. Needless to say that such a framework has been particularly useful for 
the investigation of the quantum to classical transition \cite{zeh96} as well as the phenomenon of decoherence 
\cite{cald85,milb85}.

Nevertheless, the interaction with environments having a small number of degrees of 
freedom may also cause considerable disturbances to the evolution of quantum systems. 
An interesting study in this respect is presented in \cite{akoury07}, where it is shown that even a single electron, constituting a 
``minimal environment" is enough to affect the interference fringes of another electron (system of interest) in an experiment of double 
photoionization of $\mbox{H}_2$ molecules.
In \cite{hanggi09} it is discussed the behaviour of the specific heat of quantum systems in contact with an environment 
containing just a single oscillator; the authors conclude that such a simple model is very useful to clarify the 
occurrence of anomalous effects related to the specific heat of simple systems. In another work \cite{avb14}, it is shown that a 
very small (but noisy) environment interacting with a bipartite (qubit-oscillator) system may lead to an irreversible-like 
behaviour. In summary, the above mentioned works show that even minimal environments 
might be able to cause a considerable degradation of the quantum properties of a system. 

The past years have witnessed important developments regarding the manipulation of individual quantum systems, e.g., quantum 
nanomechanical (or micromechanical) oscillators \cite{milb11,aspel14}. We may cite, for instance,
the cooling of mechanical oscillators to their quantum mechanical ground states \cite{cleland10,teufel11,chan11} and the quantum 
squeezing of motional degrees of freedom in an optomechanical system \cite{schwab15}. Other examples of physical realizations and
preparation of states of quantum harmonic oscillators are trapped ions systems \cite{wine96} and also one mode of the electromagnetic 
cavity field \cite{haro92}. A quantum oscillator may be in principle prepared in a variety of quantum states. A pure state that 
stands out is the coherent state, the ``quasiclassical" state of the oscillator defined in 
the early days of Quantum Theory \cite{schro26} and a few decades later reintroduced by Klauder \cite{klaud60}, Glauber 
\cite{glaub63} and Sudarshan \cite{sudar63}. Coherent states of the oscillator, here represented by $|\alpha\rangle 
$\footnote{Being $\alpha$ a complex number with $\hat{a} |\alpha\rangle = \alpha |\alpha\rangle$, for an oscillator associated to
creation and annihilation operators $\hat{a}^\dagger,\hat{a}$.}, have peculiar statistical properties, e.g., 
they are minimum uncertainty states in phase space. 
Besides, they have a characteristic behaviour when in contact with external systems. 
Namely, if an oscillator initially prepared in a coherent state is assumed to be linearly coupled to a reservoir at 
$T = 0$ K, its evolution is such that $|\alpha (t)\rangle = |\alpha_0 e^{-\gamma t}\rangle$; here $\gamma$ is a decay constant 
related to the oscillator-bath coupling and $\alpha_0$ is the amplitude of the initial coherent state. 
Actually, the coherent states are the only pure states that remain pure under dissipation at zero temperature \cite{dutra98}. 
But as we are going to see, in spite of their robustness at $T = 0$ K, if the oscillator in a coherent state is put in contact 
with a thermal environment at $T\neq 0$ K, its quantum state evolves to a statistical mixture of pure states i.e., 
the coherent states are no longer ``pointer states". I would like to remark that we often find in the literature discussions
about the influence of an environment on superpositions of coherent states (Schr\"{o}dinger ``cat" states) 
\cite{milb85,oliveira99,avb92,kim92,hyuga96,tombesi97,knight97} rather than individual coherent states. Differently from 
coherent states, though, ``cat" states are highly non-classical states \cite{avb92}, and their quantum properties are normally
destroyed if they are coupled to an environment even at $T = 0$ K \cite{milb85,avb92}.
Of course, in a finite temperature environment the situation is even worse \cite{kim92,hyuga96}.

Considering the model of reservoir as being a collection of oscillators \cite{louisell73,cald85,milb85}, the smallest possible environment 
could be the one consisting of a single oscillator. Thus we would have a system constituted by two coupled quantum harmonic 
oscillators; oscillator $A$, the system of interest, and oscillator $B$, the environment. The problem of two coupled oscillators 
(e.g., position-position coupling) has been already addressed in the literature; an exact analytical solution (under the rotating wave approximation) 
was found some time ago \cite{narducci68}. More recently this configuration has been considered for investigating the information transfer 
between two subsystems \cite{oliveira99}. Here I would like to explore the influence of a noisy environment (oscillator $B$) on the dynamics of
the main system (oscillator $A$). 
In order to do so, I will consider two distinct forms for the oscillator-oscillator interaction Hamiltonian: position-position (amplitude) 
and cross-Kerr (phase) couplings. Oscillator $A$ will be assumed to be initially prepared in a pure coherent state 
$|\alpha_0\rangle_a$, while oscillator $B$, the minimal environment, will be in a thermal state, a state of maximum mixture for a 
fixed energy. My analysis will be based on the time-evolution of the linear entropy, $\zeta(t) = 1 - Tr \hat{\rho}_A^2(t)$, where $\hat{\rho}_A(t)$ 
is the reduced density operator of oscillator $A$, obtained by tracing over the variables of system $B$, i.e., $\hat{\rho}_A(t) = Tr_B \hat{\rho}(t)$; here
$\hat{\rho}(t)$ is the joint density operator. 
The linear entropy equals zero for a pure state and it is larger than zero for a mixed state. It is therefore a very useful function to quantify the 
degree of mixture of the quantum state of oscillator $A$. 
Evidently in the realm of simple models involving the coupling of an oscillator to a single subsystem, as described above, the evolution of the 
oscillator will have finite recurrence times, and thus a full irreversible process is not accounted for by those models. Nonetheless, simple 
analytically solvable models may be useful to gain insights into the general properties of quantum systems. 
Here, I would like to address the following questions: to what extent simple models (restricted to a sufficiently short time-scale) 
are able to mimic the decoherence process, compared to a master equation approach? Are they able to appropriately describe the influence of 
temperature on the dynamics of a quantum oscillator?

This paper is organized as follows: in section 2 I will present the analytical solutions of the models of system-environment 
interaction. In section 3 I will discuss the evolution of the linear entropy of oscillator $A$. 
In section 4 I will summarize the conclusions.

\section{Models of environment}

\subsection{Master equation approach}

Firstly I am going to consider the usual (position-position) model of system-environment interaction based on the coupling of a harmonic oscillator 
to a thermal bath constituted by a collection of uncoupled harmonic oscillators. The oscillator $A$, of frequency $\omega$, is associated with creation 
(annihilation) operators $\hat{a}^\dagger\,(\hat{a})$, and the oscillators constituting the bath with operators $\hat{b}_i^\dagger\,(\hat{b}_i)$.
The Hamiltonian of the whole system in case of an amplitude coupling, under the rotating wave approximation, may be written as 
\begin{equation}
\hat{H} = \hbar\omega \hat{a}^\dagger\hat{a} + \hbar\sum_i \omega_i \hat{b_i}^\dagger\hat{b_i} + \hbar\sum_i (g_i \hat{a}^\dagger\hat{b}_i 
+ g^*_i \hat{b}_i^\dagger\hat{a}),\label{hamilbath}
\end{equation}
where the first two terms correspond to energy of the free oscillator $A$ plus the energy of the bath, and the third term is the interaction energy with coupling
constants $g_i$. Using a perturbative standard procedure \cite{louisell73} under the Born-Markov approximation, we may derive the (well-known) master equation 
for the oscillator $A$ reduced density operator ($\hat{\rho}_A$), in the interaction picture
\begin{equation}
\frac{d\hat{\rho}_A}{dt}=\gamma(1+\overline{n})\left(2\hat{a} \hat{\rho}_A\hat{a}^\dagger  - \hat{a}^\dagger \hat{a} \hat{\rho}_A
- \hat{\rho}_A\hat{a}^\dagger \hat{a} \right) + \gamma\overline{n}\left(2\hat{a}^\dagger \hat{\rho}_A\hat{a}  - \hat{a} \hat{a}^\dagger \hat{\rho}_A
- \hat{\rho}_A\hat{a} \hat{a}^\dagger \right).\label{mastereq}
\end{equation}
The parameter $\gamma$ is a decay constant related to the couplings $g_i$, and $\overline{n}$ is the average number of excitations 
of the oscillator of frequency $\omega$ at thermal equilibrium,
\begin{equation}
\overline{n} = \frac{1}{\exp(\hbar\omega/k_B T)-1}.
\end{equation}
A trace over the bath variables has already been taken. The system of interest (oscillator $A$) will be assumed to be initially 
prepared in a coherent state 
$|\psi(0)\rangle_a = |\alpha_0\rangle_a$. The solution of master equation (\ref{mastereq}), which is the time-evolved density operator
of oscillator $A$, may be cast in the following operator form \cite{hyuga96}
\begin{equation}
\hat{\rho}_A(t) = \sum_k \frac{\overline{n}_t^k}{(1 + \overline{n}_t)^{(k+1)}}\hat{D}(\alpha_t) |k\rangle_a \, 
{}_a\langle k|\hat{D}^\dagger(\alpha_t),
\end{equation}
where $\hat{D}(\alpha_t) = \exp(\alpha_t\hat{a}^\dagger - \alpha_t^*\hat{a})$ is Glauber's displacement operator, with 
$\alpha_t = \exp(-\gamma t)\alpha_0$ and $\overline{n}_t = [1 - \exp(-2\gamma t)]\overline{n}$. In other words, the quantum 
state of oscillator $A$ may be written as a statistical mixture of 
displaced number states with a (time-dependent) thermal distribution weight. Of course for long times, the state of
oscillator $A$ becomes the thermal equilibrium state, i.e., 
$\hat{\rho}_A = \sum_k \frac{\overline{n}^k}{(1 + \overline{n})^{(k+1)}}|k\rangle_a \, {}_a\langle k|$.
From $\hat{\rho}_A(t)$, one may calculate the time-evolution of the oscillator's 
linear entropy, $\zeta_1(t) = 1 - Tr\hat{\rho}_A^2(t)$, which reads
\begin{equation}
\zeta_1(t) = 1 - \frac{1}{1+2\overline{n}\left( 1 - e^{-2\gamma t}\right)}.\label{linentres}
\end{equation}
The linear entropy in this case is a simple function of $\gamma$ and $\overline{n}$, and does not depend on the amplitude 
$\alpha_0$ of the initial coherent state. 
The quantum system is assumed to be coupled to a very large and immutable reservoir, 
which naturally leads to irreversible decoherence. 
Thus, for long enough times, the linear entropy of oscillator $A$ tends towards a constant (maximum) value 
$\zeta_1^{max} = 2\overline{n}/(1 + 2\overline{n})$, meaning that it will eventually reach thermal equilibrium with the reservoir. 
If $\overline{n} = 0$ (reservoir at $T = 0$ K), we obtain $\zeta_1(t) = 0$ as expected. In this case the
state of oscillator $A$ remains pure at all times, i.e., $\hat{\rho}_A(t) \equiv |\alpha_0\rangle_a{}_a\langle\alpha_0 |$. 

\subsection{Amplitude coupling}

A natural example of minimal environment would be the extreme case in which the reservoir discussed in the former section, 
Eq. (\ref{hamilbath}) is reduced to a single sub-system, namely oscillator $B$ with operators $\hat{b},\hat{b}^\dagger$.  
The interaction Hamiltonian then reads
\begin{equation}
\hat{H}_I = \hbar\kappa(\hat{a}^\dagger\hat{b} + \hat{b}^\dagger\hat{a}),
\end{equation}
being $\kappa$ the coupling constant. The natural frequencies of oscillators $A$ and
$B$ are assumed to be equal, $\omega_a = \omega_b = \omega$.
An analytical solution to this problem can be found in \cite{narducci68}.
The Heisenberg equations of motion $i\hbar\frac{d\hat{O}}{dt} = [\hat{O},\hat{H}]$ for the operators $\hat{a}$ and $\hat{b}$ are
\begin{eqnarray}
\frac{d\hat{a}}{dt} &=& -i\omega\hat{a} -i\kappa\hat{b} \nonumber\\ 
\frac{d\hat{b}}{dt} &=& -i\omega\hat{b} -i\kappa\hat{a},\label{heiseneq}
\end{eqnarray}
with corresponding solutions
\begin{eqnarray}
\hat{a}(t) &=& A(t)\hat{a}(0) + B(t)\hat{b}(0) \nonumber\\ 
\hat{b}(t) &=& B(t)\hat{a}(0) + A(t)\hat{b}(0),\label{solheiseneq}
\end{eqnarray}
where $A(t) = \exp(-i\omega t )\cos\kappa t$ and $B(t) = -i\exp(-i\omega t )\sin\kappa t$.
Following \cite{narducci68}, we may then calculate Glauber's $P$-function \cite{glaub63}, $P_A(\alpha)$, 
a possible representation of the quantum state in the coherent 
state basis. It is related to the density operator via the integral in phase-space
\begin{equation}
\hat{\rho}_A = \int_{-\infty}^\infty d^2 \alpha \,P_A(\alpha) |\alpha\rangle\langle\alpha |,
\end{equation}
where $|\alpha\rangle$ are coherent states and $d^2 \alpha \equiv d({\mbox{Re}}\,\alpha) d({\mbox{Im}}\,\alpha)$. 
The $P$-function may be expressed in terms of the normally-ordered characteristic function $\chi_{NA}$ \cite{glaub63,narducci68} as
\begin{equation}
P_A = \frac{1}{\pi^2}\int_{-\infty}^\infty d^2 \eta \,\chi_{NA} e^{-\eta\alpha^* + \eta^*\alpha},\label{pfunccarac}
\end{equation}
being $\chi_{NA}$ associated to a density operator $\hat{\rho}$,
\begin{equation}
\chi_{NA}(\eta,t) = Tr\left[ \hat{\rho}(t)e^{\eta\hat{a}^\dagger} e^{-\eta^*\hat{a}}\right] = Tr\left[ \hat{\rho}(0)e^{\eta\hat{a}^\dagger(t)} e^{-\eta^*\hat{a}(t)}\right].\label{carac}
\end{equation}
Now the solutions in Eq. (\ref{solheiseneq}), $\hat{a}^\dagger(t)$ and $\hat{a}(t)$, may be inserted in the corresponding right-hand side of 
Eq. (\ref{carac}). We consider the oscillator $A$ initially
prepared in the coherent state $\hat{\rho}_A(0) = |\alpha_0\rangle_a{}_a\langle\alpha_0 |$ and oscillator $B$ (minimal environment), initially in the thermal state
\begin{equation}
\hat{\rho}_B(0) = \sum_{k=0}^\infty \frac{\overline{n}^k}{(1 + \overline{n})^{k+1}} |k\rangle_b{}_b\langle k|.\label{thermalb}
\end{equation} 
After performing the trace and integrating, one obtains the $P$-function \cite{narducci68}
\begin{equation}
P_A(\alpha,t) = \frac{1}{\pi S(t)} \exp{\left[-\frac{\left|\alpha - C(t)\right|^2}{S(t)}\right]},
\end{equation}
where $S(t) = \overline{n}\sin^2\kappa t$ and $C(t) = \alpha_0 e^{-i\omega t}\cos\kappa t$. Hence, the resulting 
$P$-function is a Gaussian function with variable width.

Now we may calculate the linear entropy of oscillator $A$, or
\begin{equation}
\zeta_2 = 1 - \frac{1}{\pi}\int_{-\infty}^\infty \int_{-\infty}^\infty \int_{-\infty}^\infty d^2 \gamma d^2 \alpha d^2 \beta
P_A(\alpha,t) P_A(\beta,t) \,\langle\gamma |\alpha\rangle \langle\alpha |\beta\rangle \langle\beta |\gamma\rangle.
\end{equation}
Using $\langle\alpha |\beta\rangle = \exp(-|\alpha|^2/2 -|\beta|^2/2 + \alpha^*\beta)$ and integrating, we obtain the linear entropy
\begin{equation}
\zeta_2(t) = 1 - \frac{1}{1+2\overline{n}\sin^2\kappa t},\label{linentamp}
\end{equation}
which is a simple function of time. At times $t_{r;m}^{(2)} = m\pi/\kappa$ $(m = 1,\, 2\ldots)$, the state of oscillator $A$ returns to 
its initial (pure) state. Yet, similarly to the linear entropy calculated from the master equation [see Eq. (\ref{linentres})], 
$\zeta_2$ does not depend on $\alpha_0$ and its maximum is $\zeta_2^{max} = \zeta_1^{max} = 2\overline{n}/(1 + 2\overline{n})$. 
Accordingly, if $\overline{n} = 0$ we have $\zeta_2(t) = 0$.

\subsection{Phase coupling}

I would like now to consider another type of coupling to a minimal environment based on a cross-Kerr interaction, or
\begin{equation}
\hat{H}_I = \hbar\lambda\hat{a}^\dagger\hat{a}\hat{b}^\dagger\hat{b},\label{hamkerr}
\end{equation}
being $\lambda$ the (cross-Kerr) coupling constant. Again, I am assuming $\omega_a = \omega_b = \omega$.
In this case, because of the form of $\hat{H}_I$ in Eq. (\ref{hamkerr}), there will 
be no energy exchange between oscillator $A$ and oscillator $B$, as occurs in the previous case.
For initial coherent ($|\alpha_0\rangle_a$) and Fock ($|m\rangle_b$) states , the evolution according to Hamiltonian (\ref{hamkerr}) will result
\begin{eqnarray}
|\Psi(t)\rangle &=& e^{-i\hat{H}_I t/\hbar} |\alpha_0\rangle_a \otimes |m\rangle_b = \sum_{n=0}^\infty \frac{e^{-\frac{|\alpha_0|^2}{2}}\alpha_0^n}{n!}
e^{-i\lambda\hat{a}^\dagger\hat{a}\hat{b}^\dagger\hat{b}}|n\rangle_a |m\rangle_b = \\ \nonumber
&=& \sum_{n=0}^\infty \frac{e^{-\frac{|\alpha_0|^2}{2}}(\alpha_0e^{-i\lambda m t})^n}{n!} |n\rangle_a |m\rangle_b.
\end{eqnarray}
Therefore, the time evolution for the joint (two oscillators system) density operator having as initial states the coherent state 
$\hat{\rho}_A(0) = |\alpha_0\rangle_a{}_a\langle \alpha_0 |$ for oscillator $A$ and the thermal state in Eq. (\ref{thermalb}) for the minimal environment 
will be 
\begin{equation}
\hat{\rho}(t) = e^{-i\hat{H}_I t/\hbar} \hat{\rho}_A(0)\otimes\hat{\rho}_B(0) e^{i\hat{H}_I t/\hbar} =  
\sum_{k=0}^{\infty} \frac{\overline{n}^k}{(1 + \overline{n})^{k+1}} \, |\alpha_k(t)\rangle_a \, 
|k\rangle_b{}_b\langle k|\, {}_a\langle \alpha_k(t)|,
\end{equation}
with $\alpha_k (t) = \alpha_0 e^{-i\lambda k t}$.
Now we may take the trace over oscillator's $B$ variables, obtaining the reduced density operator of oscillator $A$, $\hat{\rho}_A(t) = Tr_B\left[ \hat{\rho}(t) \right]$,
\begin{equation}
\hat{\rho}_A(t) = \sum_{k=0}^{\infty} \frac{\overline{n}^k}{(1 + \overline{n})^{(k+1)}} |\alpha_k(t)\rangle_a{}_a\langle \alpha_k(t)|.\label{rhoaamp}
\end{equation}
As we see from Eq. (\ref{rhoaamp}), $\hat{\rho}_A(t)$ is a mixture of coherent states with a thermal distribution weight. Again, 
if $\overline{n} = 0$, i.e., for the minimal environment initially in its vacuum state, oscillator 
$A$ will remain in its initial pure state $\hat{\rho}_A(t) \equiv |\alpha_0\rangle_a{}_a\langle\alpha_0 |$. 
The linear entropy $\zeta_3(t) = 1 - Tr \hat{\rho}_A^2(t)$ reads
\begin{equation}
\zeta_3(t) = 1 - \sum_{k=0}^\infty \sum_{l=0}^\infty \frac{\overline{n}^k}{(1 + \overline{n})^{(k+1)}} 
\frac{\overline{n}^l}{(1 + \overline{n})^{(l+1)}}e^{-2|\alpha_0|^2\left[1 - \cos \{\lambda(k-l)t\}\right]},\label{linentaphase}
\end{equation}
which is also a periodic function of time. In this model, system $A$ returns to its initial state at times 
$t_{r;m}^{(3)} = 2m\pi/\lambda$ $(m = 1,\, 2\ldots)$.

\section{Discussion of the results: Linear entropy}

In order to discuss the decoherence process, I would like first to make some considerations about the time scales involved. 
The behaviour of the linear entropies given by Eqs. (\ref{linentres}), (\ref{linentamp}) and (\ref{linentaphase}) may be compared 
in a straightforward way if we make $\gamma = \kappa = \lambda$. It would be then convenient to plot the quantities $\zeta_i$
as a function of the scaled time $\gamma t$, being $\gamma$ the decay constant in the master equation approach. The reversible
models have recurrence times at $t_r^{(2)} = \pi/\kappa$ (amplitude coupling model) and 
$t_r^{(3)} = 2\pi/\lambda$ (phase coupling model). At those times the oscillator $A$ returns to its initial state 
$|\alpha_0\rangle_a$, i.e., the joint state of the system becomes separable again. The patterns of course periodically repeat. 
I recall that if the environment is at $T = 0$ K ($\overline{n} = 0$), the oscillator $A$ will continue in 
a pure state, or $\zeta_1(t) = \zeta_2(t) = \zeta_3(t) = 0$. However, we expect that a noisy environment ($\overline{n}\neq 0$) 
will have an important influence on the behaviour of oscillator $A$ and that an initially pure state will evolve to a mixed state. 
From the expressions (\ref{linentres}), (\ref{linentamp}) and (\ref{linentaphase}) we may obtain good estimates of the 
``decoherence times" as a function of $\overline{n}$ for each model
\begin{equation}
t_d^{(1)}\approx\frac{1}{4\gamma\overline{n}}, \ \ \ \ \ \ \ t_d^{(2)}\approx\frac{1}{\kappa\sqrt{2\overline{n}}},
\ \ \ \ \ \ \ t_d^{(3)}\approx\frac{1}{5\lambda\sqrt{|\alpha_0|^2\overline{n}}}.\label{dectimes}
\end{equation}
Note that $t_d^{(1)}\propto 1/\overline{n}$ for the master equation approach, while in the 
simplified models, $t_d^{(2,3)}\propto 1/\sqrt{\overline{n}}$, instead. This significant difference is related to the fact 
that in the master equation approach the bath is composed by a large number of oscillators, while in the simplified 
models the environment is reduced to a single oscillator.
I would also like to remark that some differences that may be observed in the obtained results are related to the inter-oscillator 
couplings, as well as to the fact that the initial state of oscillator $A$ is a coherent state. As a matter of fact, in both 
the master equation and amplitude coupling approaches, the interaction Hamiltonian has basically the same form, viz. 
$\hat{H}_I\propto\hat{a}\hat{b}^\dagger + c.c.$.
In this case there is energy exchange between the oscillators, e.g., a quantum of energy of oscillator $A$ is destroyed by $\hat{a}$.
At the same time, this makes a coherent state a special state, given that $|\alpha_0\rangle$ is eigenstate of $\hat{a}$.
Moreover, because the structure of the master equation is also related to the form of this coupling \cite{zoller96,davidov01}, 
an oscillator initially in a coherent state $|\alpha_0\rangle$ (eigenstate of $\hat{a}$) remains a pure state during the evolution 
if the bath is at $T = 0$ K. Besides, even if $T \neq 0$ K, the linear entropy of 
oscillator $A$ will still not depend on $\alpha_0$. On the other hand, in the phase coupling model with 
interaction $\hat{H}_I\propto\hat{a}^\dagger\hat{a}\hat{b}^\dagger\hat{b}$, we expect a dependence on $\alpha_0$, 
as the initial coherent state is not an eigenstate of $\hat{a}^\dagger\hat{a}$.

\begin{figure}[h]
	\resizebox{0.7\columnwidth}{!}{
		\includegraphics{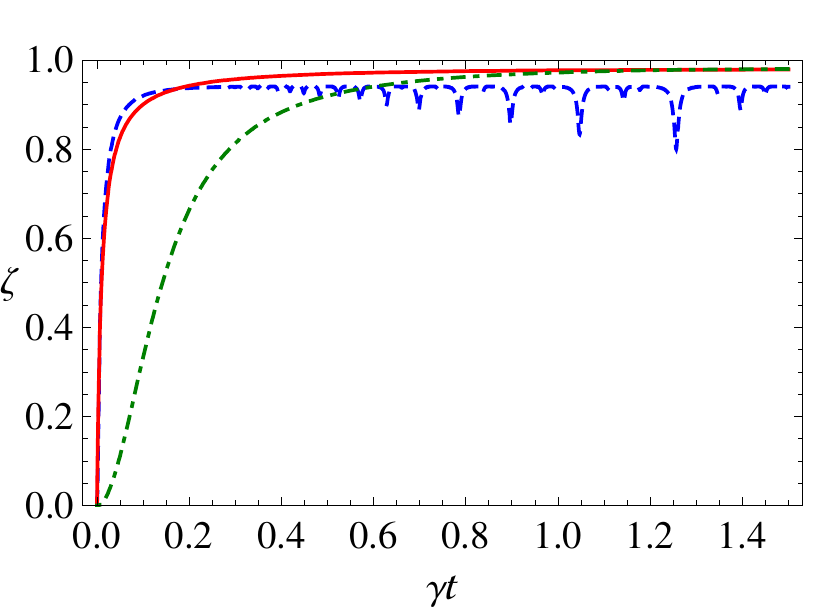}}
	\caption{\label{figure1} Linear entropies as a function of the scaled time $\gamma t$ ($\gamma = \kappa = \lambda$) for: 
	a) the master equation model ($\zeta_1$; red, continuous line);
	b) the amplitude coupling model ($\zeta_2$; green, dot-dashed line), and c) the phase coupling model ($\zeta_3$; blue, dashed line).
	Here $\alpha_0 = 5$, $\overline{n} = 25$.}
\end{figure}

\begin{figure}[h]
	\resizebox{0.7\columnwidth}{!}{
		\includegraphics{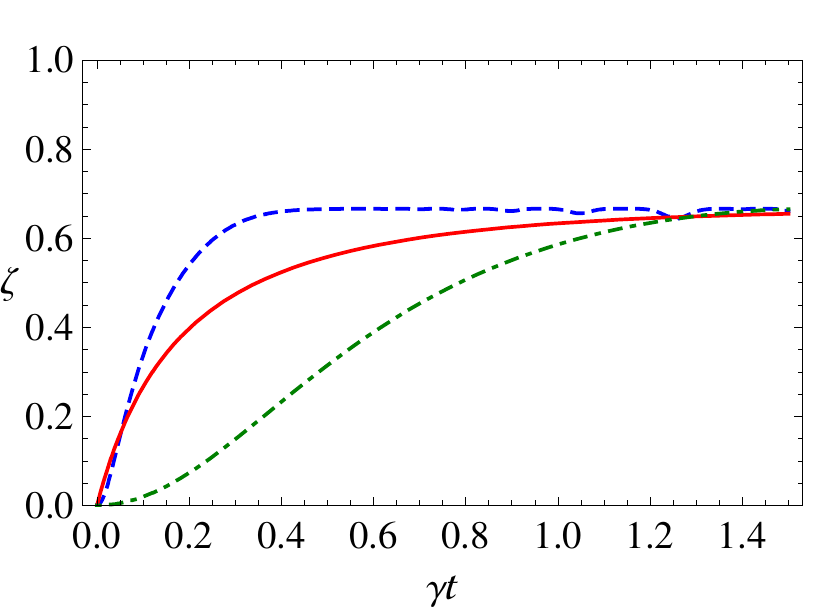}}
	\caption{\label{figure2} Linear entropies as a function of the scaled time $\gamma t$ ($\gamma = \kappa = \lambda$) for: 
	a) the master equation model ($\zeta_1$; red, continuous line);
	b) the amplitude coupling model ($\zeta_2$; green, dot-dashed line), and c) the phase coupling model ($\zeta_3$; blue, dashed line).
	Here $\alpha_0 = 5$, $\overline{n} = 1$.}
\end{figure}

\begin{figure}[h]
	\resizebox{0.7\columnwidth}{!}{
		\includegraphics{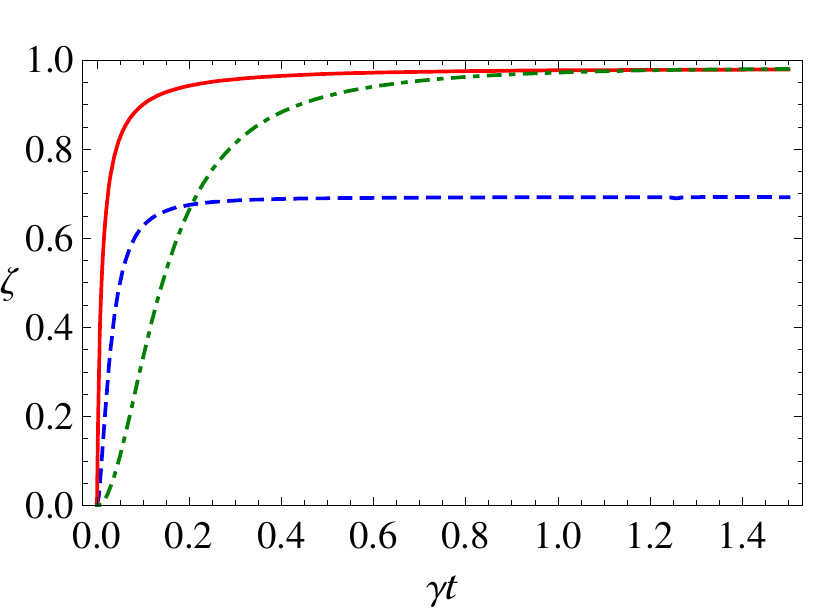}}
	\caption{\label{figure3} Linear entropies as a function of the scaled time $\gamma t$ ($\gamma = \kappa = \lambda$) for: 
	a) the master equation model ($\zeta_1$; red, continuous line);
	b) the amplitude coupling model ($\zeta_2$; green, dot-dashed line), and c) the phase coupling model ($\zeta_3$; blue, dashed line).
	Here $\alpha_0 = 1$, $\overline{n} = 25$.}
\end{figure}

Now we may proceed with a graphical analysis of the linear entropies in order to give a clearer picture of the 
decoherence process. In Fig. (\ref{figure1}) we have plots of the linear entropies as a function of time, for oscillator $A$ 
initially prepared in a coherent state with $\alpha_0 = 5$, and the mean excitation number associated to the environment,  
$\overline{n} = 25$. We immediately observe that all three curves have a similar general behaviour; the quantum state of oscillator $A$, 
initially pure, rapidly becomes a statistical mixture, and the linear entropies reach a plateau. We also note some differences, such as
oscillations (dips) in the curve obtained from the phase coupling model, which are due to the terms having different frequencies in Eq. 
(\ref{linentaphase}). Furthermore, decoherence is significantly slower in the amplitude coupling model in contrast to the
master equation and phase coupling models. This is consistent with the estimated decoherence times in Eqs. (\ref{dectimes});
if $\overline{n}$ is comparable to $ |\alpha_0|^2$, we have a reasonably good agreement between the curves of $\zeta_1$ and $\zeta_3$ 
as shown in Fig. (\ref{figure1}). In this case it is verified a strong decoherence, given that the amount of noise in 
the environment is relatively high ($\overline{n} = 25$). Nevertheless, the initial coherent state of oscillator $A$ will 
evolve to a statistical mixture even for modest values of $\overline{n}$. In Fig. (\ref{figure2}) the linear entropies 
are plotted as a function of time for a considerably lower temperature of the environment, or $\overline{n} = 1$, but still 
having $\alpha_0 = 5$. Firstly we note that the maximum values of the linear entropy decrease with decreasing temperatures 
for each model, even though the slopes of the curves differ considerably. In this particular case $\zeta_3$ (phase coupling model) 
reaches its maximum value faster than the linear entropies given by the other two models. 
In fact for $\alpha_0 = 5$ and $\overline{n} = 1$, the decoherence time $t_d^{(3)}$ is actually shorter than the others 
[see Eqs. (\ref{dectimes})]. Another difference concerning the phase coupling model, 
is that the maximum value $\zeta_3^{max}$ also decreases with decreasing values of $\alpha_0$, while 
in both the master equation approach and the amplitude coupling model, the corresponding maximum values 
of the linear entropies depend only on $\overline{n}$, or $\zeta_1^{max} = \zeta_2^{max} = 2\overline{n}/(1 + 2\overline{n})$. 
Thus, for smaller $\alpha_0$, the linear entropy $\zeta_3$ may not match the plateaus of $\zeta_1$ and 
$\zeta_2$, as shown in Fig. (\ref{figure3}), with $\alpha_0 = 1$ and $\overline{n}=25$. In the master equation model, 
the plateau corresponds to the (thermal) steady state, clearly characterizing an irreversible behaviour. 
Yet in the case of the simple (reversible) models, the recurrence times are finite, and 
oscillator $A$ returns to is original state. Nonetheless, the plateau representing a mixed state may survive for times 
much longer than the typical decoherence times; for instance, as seen in Fig. (\ref{figure1}),
$t_d^{(3)}\approx 1/5\lambda\sqrt{|\alpha_0|^2\overline{n}} \approx 0.008/\lambda\ll t_r^{(3)}$. Even for smaller $\overline{n}$,
as in Fig. (\ref{figure2}), we may have relatively short decoherence times, i.e., $t_d^{(3)}\approx 0.2/\lambda$.

\begin{figure}[h]
	\resizebox{0.7\columnwidth}{!}{
		\includegraphics{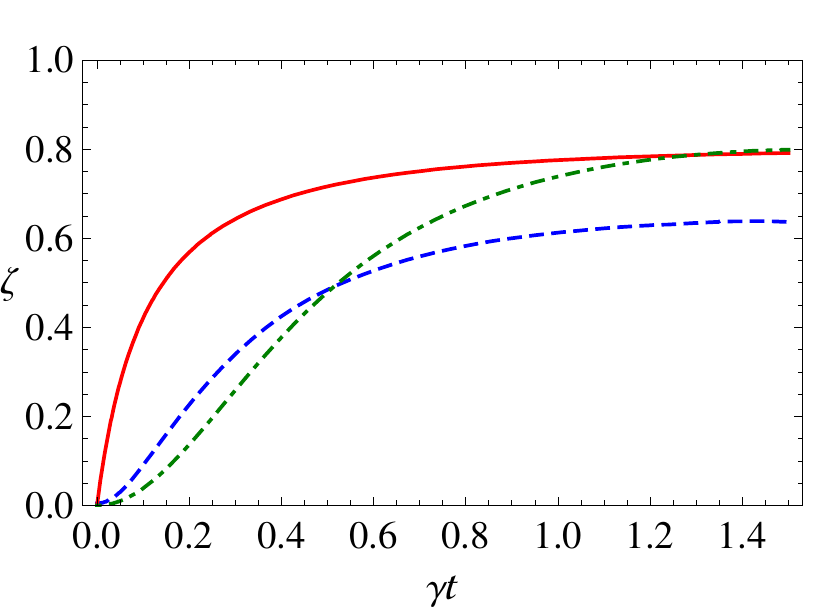}}
	\caption{\label{figure4} Linear entropies as a function of the scaled time $\gamma t$ ($\gamma = \kappa = \lambda$) for: 
	a) the master equation model ($\zeta_1$; red, continuous line);
	b) the amplitude coupling model ($\zeta_2$; green, dot-dashed line), and c) the phase coupling model ($\zeta_3$; blue, dashed line).
	Here $\alpha_0 = 1$, $\overline{n} = 2$.}
\end{figure}

\begin{figure}[h]
	\resizebox{0.7\columnwidth}{!}{
		\includegraphics{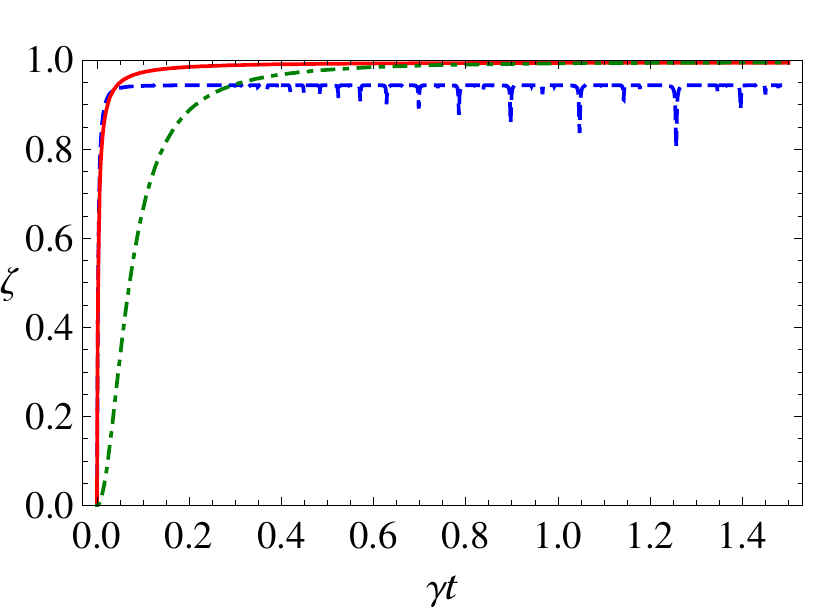}}
	\caption{\label{figure5} Linear entropies as a function of the scaled time $\gamma t$ ($\gamma = \kappa = \lambda$) for: 
	a) the master equation model ($\zeta_1$; red, continuous line);
	b) the amplitude coupling model ($\zeta_2$; green, dot-dashed line), and c) the phase coupling model ($\zeta_3$; blue, dashed line).
	Here $\alpha_0 = 5$, $\overline{n} = 100$.}
\end{figure}

In the previous examples it seems that the phase coupling model, rather than the amplitude coupling model, is the one which 
has a short time dynamics (at least qualitatively) more similar to the one obtained from the master equation model. However, this may 
change depending on the values of the parameters involved, e.g., for smaller $\alpha_0$, as shown in Fig. (\ref{figure4}), with 
$\overline{n} = 2$ and $\alpha_0 = 1$, the phase coupling model curve is clearly closer to the curve obtained from the
amplitude coupling model. On the other hand, for very large values of $\overline{n}$, the linear entropies according to the master equation 
and phase coupling models may almost coincide (for short times), as shown in Fig. (\ref{figure5}), with $\overline{n} = 100$ 
and $\alpha_0 = 5$.

\section{Conclusions}
 
I have presented a study of the dynamics of a quantum oscillator prepared in a quasi-classical (coherent) state in interaction 
with a very small bath constituted by a single oscillator which is initially in a thermal state. I have considered two types of 
inter-oscillator couplings, and a comparison has been made with the evolution of the linear entropy obtained via a widely 
used model of open quantum systems, namely the master equation approach. 
In the amplitude and phase coupling models for the small environments here contemplated, we have an intrinsic non-Markovian behaviour, 
reversible dynamics and exact solutions, while in the master equation model (large reservoir) the evolution is Markovian, irreversible 
and the solution is approximate (perturbative). Despite being very different approaches, if times are short enough, the linear entropies 
obtained from each model may have very similar features. It is possible, within the realm of the simple models, to emulate some 
characteristic features of the decoherence process as described by the master equation approach, such as the fast evolution of oscillator $A$
(initially in a pure state) towards a statistical mixture. Besides, given that the oscillators become entangled during most of the time, 
oscillator $A$ may also spend a relatively long time in a mixed state. Of course it was not the aim here to give a comprehensive description of the 
phenomenon of decoherence, but rather, to identify common features as well as differences between the master equation approach and simple 
models. I believe that this study may contribute to a better understanding of the decoherence process itself, as well as to the investigation 
of the behaviour of simple quantum systems embedded in few-body environments.



\section*{Acknowledgements}

I would like to thank CNPq (Conselho Nacional para o 
Desenvolvimento Cient\'\i fico e Tecnol\'ogico) and FAPESP
(Funda\c c\~ao de Amparo \`a Pesquisa do Estado de S\~ao Paulo), Brazil,
for financial support through the National Institute for Science and 
Technology of Quantum Information (INCT-IQ) and the Optics and Photonics 
Research Center (CePOF)

\newpage

\end{document}